\documentclass[12pt]{article}

\usepackage{amssymb}
\usepackage{amscd}
\usepackage{graphics}
\usepackage{amsfonts}
\usepackage{amsmath}
\usepackage{latexsym}
\usepackage{color}
\usepackage{empheq}
  \usepackage{tikz}
\usepackage{tikz-3dplot}
\usetikzlibrary{calc,3d,arrows}
{\left\lbrace\begin{array}{@{}l@{}}}%
{\end{array}\right.}

\usepackage{amsthm}
\usepackage{authblk}

\usepackage{graphicx}% Include figure files
\usepackage{dcolumn}% Align table columns on decimal point
\usepackage{bm}% bold math
\usepackage[utf8]{inputenc}
\usepackage[T1]{fontenc}
\usepackage{etoolbox}

\def\ms{\medskip}
\def\l{\left(}
\def\r{\right)}

\usepackage{latexsym,graphics,color}
\def\bes{\begin{subequations}}
\def\ees{\end{subequations}}
\def\ba{\begin{align}}
\def\ea{\end{align}}

\def\sn{\mathrm{sn}}
\def\cn{\mathrm{cn}}
\def\dn{\mathrm{dn}}
 \def\ekn{e^{-\ri\ka\nu}}

\def\js{\frac{1}{4}}
\def\dg{\dagger}

\def\be{\begin{equation}}
\def\ee{\end{equation}}
\def\D{\mathcal D}

\def\K{\mathcal K}

\def\E{\mathcal E}
\def\M{\mathcal M}
\def\P{\mathcal P}
\def\Q{\mathcal Q}

\def\B{\mathcal B}

\def\bc{\mathbb C}
\def\br{\mathbb R}

\def\bs{\boldsymbol}
\def\k{\kappa}
\def\s1{\sigma^1}
\def\s2{\sigma^2}
\def\s3{\sigma^3}

\def\dg{\dagger}
\def\noi{\noindent}

\def\T{{\cal T}}
\def\si{\sigma}

\def\d{\partial}
\def\jp{\frac{1}{2}}
\def\ri{{\mathrm i}}

\def\Ad{{\rm Ad}}
\def\ad{{\rm ad}}

\def\epk{e^{\ka E}}
\def\emk{e^{-\ka E}}
\def\epkd{e^{\ka E^\dg}}
\def\emkd{e^{-\ka E^\dg}}
\definecolor{lila}{rgb}{1,0.2,0.9}
\definecolor{brown}{rgb}{0.5,0.3,0.3}
\definecolor{turquoise}{rgb}{0.2,0.9,0.7}
\definecolor{Orange}{rgb}{0.93,0.44,0}           %238,112,1 /255
\definecolor{GrayBlue}{rgb}{0.35,0.4,0.62}       %13,50,69 /255
\definecolor{SeafoamGreen}{rgb}{0.54,0.71,0.50}  %137,180,128 /255

\definecolor{darkorange}{cmyk}{.20,.50,.80,0}
\definecolor{lightorange}{cmyk}{.07,.37,.65,0}
\definecolor{darkpeagreen}{cmyk}{.50,.30,.50,0}
\definecolor{lightpeagreen}{cmyk}{.22,.20,.40,0}

\theoremstyle{definition}

\theoremstyle{definition}

\theoremstyle{definition}

\def\no{\noindent}                      %
\def\G{{\cal G}}

\def\ri{{\mathrm{i}}}                   %
\def\1{{\mbox{\boldmath $1$}}}          %
\def\tr{\mathrm{tr\,}}                  %
\def\e{\epsilon}  %
\def\ka{\kappa}
\def\lm{\lambda}                        %
\def\bt{\beta}                          %
\def\jp{\frac{1}{2}}                    %
\def\om{\omega}                         %
\def\al{\alpha}        
\def\bs{\boldsymbol}%

\definecolor{spec}{rgb}{0.0, 0.26, 0.15}

\def\bpm{\begin{pmatrix}}
\def\epm{\end{pmatrix}}

\DeclareMathSymbol{\Rho}{\mathalpha}{operators}{"50}

\begin{document}
 
\begin{flushright}
{}~
  
\end{flushright}

\vspace{1cm}
\begin{center}
{\large \bf On integrable deformations of the Cherednik model}

\vspace{1cm}

{\small
{\bf Ctirad Klim\v{c}\'{\i}k}
\\
Aix Marseille Universit\'e, CNRS, Centrale Marseille\\ I2M, UMR 7373\\ 13453 Marseille, France}
\end{center}

\vspace{0.5 cm}

\centerline{\bf Abstract}
\vspace{0.5 cm}
\noindent   We provide  the $\E$-model formulation  of the non-deformed Cherednik model as well as of its  Poisson-Lie and Poisson-Lie-WZ deformed version.  In all three cases we 
solve the sufficient   condition of integrability by using the $\E$-model formalism.
We thus recover in an alternative way the known results for the non-deformed and the Poisson-Lie deformed models, while for the Poisson-Lie-WZ deformed one our
results are new.

%%%%%%%%%%%%%%%%%%%%%%%%%%%%%
\section{Introduction} In the present paper, we study the integrability of
three families of
$(1+1)$-dimensional $\sigma$-models living on a simple compact group $K$.
 Here are the respective actions of those three families
  \be S=\jp\int d\tau  d\si \l E^{-1}\d_+kk^{-1}, \d_-kk^{-1}\r_\K, \label{271}\ee
    \be S_\nu(k)=\jp\int d\tau  d\si \l (E+\nu R_{k^{-1}})^{-1}\d_+kk^{-1}, \d_-kk^{-1}\r_\K, \label{290b}\ee
         \begin{multline} S_{\ka,\nu}(k)=\frac{\ka}{4}\int d\tau d\si  \biggl( \frac{1+e^{-\ka\nu R_{k^{-1}}}e^{-\ka E }}{1-e^{-\ka\nu R_{k^{-1}}}e^{-\ka E }}\partial_+ kk^{-1},\partial_- k k^{-1}\biggr)_\K +\\+\frac{\ka}{4}\int d^{-1}\oint (dkk^{-1},[\partial_\sigma kk^{-1},dkk^{-1}])_\K.\label{293}\end{multline}
%{\color{blue}  \begin{multline} S_{\ka,\nu}(k)=\frac{\ka}{4}\int d\tau d\si  \biggl( k^{-1}\partial_+ k,\frac{1+e^{-\ka E_k^\dg}e^{\ka\nu R}}{1-e^{-\ka E_k^\dg}e^{\ka\nu R}}k^{-1}\partial_- k\biggr)_\K +\\+\frac{\ka}{4}\int d^{-1}\oint (k^{-1}dk,[k^{-1}\partial_\sigma k,k^{-1}dk])_\K.\label{293bis}\end{multline}}
Every member of each of the three families is characterized by an invertible linear operator
$E:\K\to\K$  acting on the
  Lie algebra $\K$ of $K$ . The remaining notations are as follows:
  $\tau, \si$ are the standard coordinates on the cylindrical world-sheet,
 the symbols $\d_\pm=\d_\tau\pm\d_\si$ stand for the light-cone derivatives,  $k(\tau,\si)\in  K$ is a $\sigma$-model field configuration 
 and $(.,.)_\K$ is an appropriately normalized Killing-Cartan bilinear form. The   linear operator $R:\K\to \K$ is the so-called Yang-Baxter operator to be defined in Section 4.1 and we denote $R_{k^{-1}}:=\Ad_kR\Ad_{k^{-1}}$.
Finally, it is easy to see that $\ka$ and $\nu$ are deformations parameters, that is 
  in the limit $\ka\to 0$ the family \eqref{293} becomes \eqref{290b} and in the limit 
  $\nu\to 0$ the family \eqref{290b} becomes \eqref{271}.
  
  \ms 
  
  Historically,  the non-deformed family \eqref{271} was studied already  by Zakharov and Mikhailov \cite{ZM}  and also by Cherednik \cite{C81} while the $\nu$-deformed family \eqref{290b}   was  introduced 
   in \cite{KS96} and it is referred to as the Poisson-Lie   deformation
 of  \eqref{271}. On the other hand, the family  \eqref{293} is  proposed here  for the first time, we take the liberty to call it the Poisson-Lie-WZ deformation of \eqref{271}. Note however that for the   operators $E$ of a particular form $E=\al\1+\beta R$ the model \eqref{293}
 was already studied in \cite{K20}, where it was proven that it is equivalent to the DHKM model proposed in \cite{DHKM}.

 \ms 

It turns out that for the choice $E=\al\1+\beta R$ ($\al,\bt$ are numbers and $\1$ the identity operator on $\K$), all three models \eqref{271}, \eqref{290b} and \eqref{293} are integrable, they are named respectively as  the Yang-Baxter $\si$-model,
the bi-Yang-Baxter one and the bi-Yang-Baxter-WZ one\footnote{ In the case $\bt=0$, that is $E=\al\1$, the three models are respectively called   the principal chiral one, the Yang-Baxter one and the Yang-Baxter-WZ one \cite{DMV15}.} and their integrability  was proven respectively in \cite{K09}, \cite{K14} and \cite{DHKM,K20}. 

 \ms

 Can we go beyond the choice $E=\al\1+\beta R$ while keeping the integrability?
 This question was answered affirmatively by Cherednik \cite{C81}  for the non-deformed model \eqref{271}, for the case of 
 $K=SU(2)$  and for the diagonal  operator $E={\rm diag} (D_1^{-1},D_2^{-1},D_3^{-1})$   in the standard
 $su(2)$ basis of Pauli matrices  $T_j=-\ri\si_j$, $j=1,2,3$. If we set
  $$\d_\pm kk^{-1}=(\d_\pm kk^{-1})_jT_j,\quad (T_j,T_k)_\K=-\tr(T_jT_k),$$
  then the action of the Cherednik model can be written as
    \begin{multline} S(k)= \int d\tau  d\si D_1(\d_+ kk^{-1})_1(\d_- kk^{-1})_1 +\\+\int d\tau d\si\l D_2(\d_+ kk^{-1})_2(\d_- kk^{-1})_2+D_3 (\d_+ kk^{-1})_3(\d_- kk^{-1})_3\r.\label{280} \end{multline}
Note that if $D_1=D_2=D_3$ the Cherednik model is just the principal chiral model.
In the case  $D_1=D_2<D_3$ the Cherednik model turns out to coincide up to a total derivative with the Yang-Baxter model for $E=\frac{\1+\sqrt{D_3/D_1-1}R}{D_3}$. What is important is that even if the Cherednik model is totally anisotropic, that is  $D_1\neq D_2\neq D_3\neq D_1$,  it  is still integrable \cite{C81,LW}. Because of the appearence of elliptic functions in the Lax operator,   the totally anisotropic Cherednik  model was referred to in \cite{LW}
as the {\it elliptic deformation} of the principal chiral model. 
The elliptic deformations of the family \eqref{271} for other groups than $SU(2)$ were then constructed and studied in \cite{LW,HRS,LW24}.

\ms 

It was recently shown in \cite{KS24} that the Poisson-Lie deformed model \eqref{290b} also admits the elliptic integrable deformation in   case of the group $K=SU(2)$.

\ms 

In the present paper we study the integrability of the three families \eqref{271},
\eqref{290b} and \eqref{293} starting from the  $\E$-model approach.
This means that we formulate each of the three models  as  the $\E$-models
on appropriate Drinfeld doubles and then we try to find out whether a sufficient condition of   $\E$-model integrability \cite{S17} can be applied.   In the case of the first two families \eqref{271} and \eqref{290b}, 
  a suitable ansatz for the $\E$-model Lax matrix completely reproduces the results
   of \cite{C81,LW,KS24} which were obtained by   non-$\E$-models methods. We find however   that in the case of the Poisson-Lie-WZ deformation \eqref{293} the situation is different. The sufficient condition of integrability does not admit solutions for arbitrary couplings $D_1,D_2,D_3$,   but it does
   admit them when at least two of the couplings $D_j$ coincide.

   \ms

   The plan of the paper is as follows: in Section 2, we review the definition
and the basic properties of the $\E$-models and in Section 3, we interpret
the Cherednik model \cite{C81,LW} as well as its Poisson-Lie  deformation \cite{KS24} in terms of appropriate $\E$-models. We also give an alternative $\E$-model derivation of the corresponding Lax pairs
and RG flows and compare them with those obtained in \cite{C81,LW,KS24}. In  Section 4, we  identify  the  $\E$-model which underlies the Poisson-Lie-WZ deformation \eqref{293} and,
for the Cherednik choice $E=D^{-1}$, we classify all possible solutions of  the conditions of integrability. We finish with conclusions and outlook.

\ms

%%%%%%%%%%%%%%%%%%%%%%%%%%%%%%
 \section{Review of  $\E$-models}
 \subsection{General story}
  %%%%%%%%%%%%%%%%%%%%%%%%%%%%%%%
 Recall that the Drinfeld double $D$ is a connected even-dimensional
 Lie group   equipped with a bi-invariant pseudo-Riemannian metric of maximally-Lorentzian (split) signature.
 This pseudo-Riemannian metric   naturally induces the non-degenerate ad-invariant symmetric bilinear form $(.,.)_\D$ defined on the Lie algebra $\D$ of $D$.
 \ms 
 
The  $\E$-model  is a first-order  dynamical system living  on  the loop group $LD$ of a   Drinfeld double $D$. Its symplectic forms
is given by an expression
  \be   \omega = -\jp\oint d\si (l^{-1}dl\stackrel{\wedge}{,}\d_\si (l^{-1}d l))_{\D}\label{312},\ee
  where   $l=l(\si)$ is an element of the loop group $LD$, $\d_\si$  denotes the  derivative  with respect to the loop parameter $\si$ and   $l^{-1}dl$ is the left-invariant Maurer-Cartan form on $LD$.  The Hamiltonian of the
  $\E$-model is given by 
 \be H_\E  = \jp\oint d\si (\d_\si ll^{-1},\E \ \! \d_\si l l^{-1})_{\D},\label{284}\ee
  where $\E:\D\to\D$ is the  $\br$-linear operator squaring to identity, symmetric with respect to the bilinear form $(.,.)_\D$ and such that the bilinear form $(.,\E.)_\D$
 on $\D$ is strictly positive definite.

\ms

 The Poisson brackets of the $\D$-valued current $j=\d_\si ll^{-1}$ read 
   \begin{multline} \{(j(\sigma_1),T_1)_\D,(j(\sigma_2),T_2)_\D\}=\\ =(j(\sigma_1),[T_1,T_2])_\D\delta(\sigma_1-\sigma_2)+(T_1,T_2)_\D\d_{\si_1}\delta(\sigma_1-\sigma_2), \quad \forall T_1,T_2\in\D.\label{294}\end{multline}
 It follows that the first order Hamiltonian equations of motion of the $\E$-model are 
  \be \frac{\d j}{\d \tau}=\{j,H_\E\}=\d_\si (\E j)+[\E j,j], \quad j:=\d_\si ll^{-1}.  \label{342}\ee
Alternatively, the first-order equations of motion can be rewritten as
$$\d_\tau ll^{-1}=\E\d_\si ll^{-1}.$$

  \ms 

   Let $B$ be a half-dimensional isotropic subgroup of $D$. The   isotropy  means that the restriction of the bilinear form $(.,.)_\D$ to the Lie
   subalgebra $\B$ vanishes.  We can then  define
  a $\si$-model action for a $D$-valued field $l(\tau,\si)$ as in \cite{KS97,K21}
  \begin{multline} 
 S_\E(l)=\frac{1}{4}\int d^{-1}\oint d\si \biggl(dll^{-1},[\partial_\sigma ll^{-1},d ll^{-1}]\biggr)_\D + \\ -\frac{1}{4} \int d\tau d\si ( \partial_+ ll^{-1},W_l(-\E) \partial_- ll^{-1} )_\D+\frac{1}{4} \int d\tau d\si (W_l(+\E)\partial_+ ll^{-1}, \partial_- ll^{-1})_\D, \label{361}\end{multline}
where $l(\tau,\sigma)\in D$ is a field configuration and $W_l(\pm\E):\D\to\D$ are the projection operators   fully characterized by their respective kernels and images  
\be   \mathrm{Ker}W_l(\pm\E)=Ad_l\B,  \quad
  \mathrm{Im}W_l(\pm\E) =(1\pm\E)\D. \label{392}\ee
 The field equations coming from the action \eqref{361} have the zero curvature form
 \be \d_+(W_l(-\E)\d_-ll^{-1})-\d_-(W_l(+\E)\d_+ll^{-1})
 -\left[W_l(+\E)\d_+ll^{-1},W_l(-\E)\d_-ll^{-1}\right]_\D=0.\label{394}\ee
  The equations  \eqref{342} and \eqref{394}
  are then  related by
  \be j=\jp W_l(+\E)\d_+ll^{-1}-\jp W_l(-\E)\d_-ll^{-1}.\ee

  The $\si$-model \eqref{361} lives  seemingly  on the Drinfeld double $D$, but actually it lives on the space of cosets $D/B$ because the action $S_\E(l)$ is invariant with respect to a
  gauge symmetry $l(\tau,\si)\to l(\tau,\si)b(\tau,\si)$. 
By the general theory of the $\E$-models \cite{KS97,K20}, the first-order Hamiltonian dynamics of the $\si$-model \eqref{361} is given precisely by the $\E$-model data  $\om,H_\E$.

 %%%%%%%%%%%%%%%%%%%%%%%%%%%%%%%%%%%%%%%%%%%%%%%%%
\subsection{Integrable  $\E$-models}
 %%%%%%%%%%%%%%%%%%%%%%%%%%%%%%%%%%%%%%%%%%%%%%%%%%%

Consider a Drinfeld double $D$, the self-adjoint  linear involution $\E:\D\to\D$  and a Lie algebra $\K$. Let 
$O(\lm):\D\to\K$ be a one-parameter family of linear maps  which verify   
 \be [O(\lm)x_+, O(\lm)x_-]_\K=O(\lm)[x_+,x_-]_\D, \quad \forall x_\pm\in (\1\pm \E)\D\label{381}.\ee
Then the corresponding  $\si$-model \eqref{361} is integrable, with the 
  Lax pair  given by  \cite{S17}
\be L_\pm(\lm)= O(\lm)W_l(\pm\E)\d_\pm ll^{-1}. \label{384}\ee
If $l$ is a solution of the field equations of the $\si$-model \eqref{361},
then it holds  for every $\lm$
$$\d_+L_-(\lm)-\d_-L_+(\lm)+[L_-(\lm),L_+(\lm)]=0.$$
This is easy to see using \eqref{392}, \eqref{394}  and \eqref{381}.

  \smallskip
   \subsection{RG flow of the $\E$-models}
   
 The $\si$-models of the form \eqref{361} are automatically one-loop renormalisable \cite{VKS}, which means that
  the one-loop quantum corrections result in the RG flow of the 
operator $\E$.  This flow
is described by an elegant formula derived in \cite{SST} (and used in an different context already in \cite{T,FR})
\be\frac{d\E}{d\mu}=\E\M\E-\1\M\1, \quad \M=[[\E,\E]]-[[\1,\1]].\label{fl}\ee
Here $[[.,.]]$ is a bilinear operation which associates to two linear operators $\P,\Q:\D\to \D$ certain linear operator $[[\P,\Q]]:\D\to \D$. This operation can be defined invariantly, but for our purposes we shall define it by picking a  basis $T_A$ of $\D$. We set 
$$[T_A, T_B]=f_{AB}^{\ \ C}T_C, \quad \eta_{AB}=(T_A, \1 T_B)_\D=(T_A, T_B)_\D,  \quad \E_{AB}=(T_A,\E T_B)_\D, \quad $$ and we define the inverse matrix $\eta^{KL}$, with the help of which we can raise the indices.
Thus we have e.g. 
$$\E^{AB}=\eta^{AC}\eta^{BD}\E_{CD}.$$
 Then $[[\P,\Q]]$ is defined via its matrix elements as 
 $$ [[\P,\Q]]^{AB}=\P^{KM}\Q^{LN}f_{KL}^{\ \ A}f_{MN}^{\ \ B}.$$
 Note in this respect, that the bracket $[[\1,\1]]$ of the identity operator with itself is a non-trivial linear operator on $\D$ given entirely by the structure of the double $\D$.

   % $\A=\vz\ot\vz^\dg$, $\B=\vz\ot\vk^\dg$
 
%%%%%%%%%%%%%%%%%%%%%%%%%%%%%%%%%%%%%%%%%%%%%%%
\section{Non-deformed Cherednik model}
\subsection{General case}
%%%%%%%%%%%%%%%%%%%%%%%%%%%%%%%%%%%%%%%%%%%%%%%
It was already known to the authors of \cite{LW} that the $\si$-models of the family \eqref{271} admit an $\E$-model formulation \cite{L}. We give here the details of the construction, that is 
we  determine  the Drinfeld double $D$, the involution $\E:\D\to\D$ and the isotropic subalgebra $\B$
in such a way that the non-deformed $\si$-model \eqref{271} can be written
in the $\E$-model form \eqref{361}.  

\ms 

As a manifold, $D=K\times \K$ is the direct product of the simple compact group $K$ with its Lie algebra $\K$, the elements of $D$ are therefore pairs
$(k,\ka)$, $k\in K$, $\ka\in\K$. 
The group  multiplication law reads
     $$ (k_1,\ka_1)(k_2,\ka_2)=(k_1k_2,\kappa_1+{\rm Ad}_{k_1}\kappa_2),$$ and the inverse element is
     $$ (k,\k)^{-1}=(k^{-1},-\Ad_{k^{-1}}\k).$$
Finally,   if $1$ stands for the unit element of $K$ and $0$ is the neutral element of $\K$   then  the unit element of $D$ is 
      $ e=(1,0)$.

      \ms 

The elements of the Lie algebra $\D$ are pairs $(\mu,\rho)$, $\mu,\rho\in\K$, the commutator in $\D$ is given by
$$ [ (\mu_1,\rho_1),(\mu_2,\rho_2)]_\D=([\mu_1,\mu_2]_\K, [\mu_1,\rho_2]_\K+[\rho_1,\mu_2]_\K)$$
and the  symmetric non-degenerate ad-invariant  bilinear form $(.,.)_\D$ on the Lie algebra $\D$ is given by
   $$ \Bigl((\mu_1,\rho_1),(\mu_2,\rho_2)\Bigr)_\D=(\mu_1,\rho_2)_\K+
   (\mu_2,\rho_1)_\K,\quad \mu_{1,2},\rho_{1,2}\in\K.$$
   The ad-invariance of $(.,.)_\D$ can be checked by using the following formula expressing the adjoint action of $D$ on $\D$
   \be \Ad_{(k,\k)}(\mu,\rho)=(\Ad_k \mu,\Ad_k \rho+\ad_\k\Ad_k\mu)\label{437}.\ee

 \medskip
 
  Let $E:\K\to\K$ be a linear operator and denote by $E^\dg$ its adjoint with respect to the Killing-Cartan form. We set 
  $$S=\jp(E+E^\dg),\quad A=\jp(E-E^\dg)$$
  and we require that $S$ is invertible. Define a linear self-adjoint involution $\E:\D\to\D$ by 
\be \E(\mu,\rho)=  -(\mu,\rho)+(ES^{-1}(E^\dg \rho+\mu),S^{-1}(E^\dg \rho+\mu)).\label{eo}\ee

\bigskip 

Finally, we take  $B=\{(1,\ka),\ka\in\K\}$ for the  half-dimensional isotropic subgroup  $B$ 
of $D$. We note that $B$ is Abelian and the Lie subalgebra $\B$ is spanned by the elements of the form $(0,\rho)$, $\rho\in\K$. 

\medskip 

Now we make explicit the action \eqref{361} for our choice of $D,\E$ and $B$.
First we fix the gauge symmetry $l\to lb$ by setting $l=(k,0)$, which has a consequence that the WZ term in \eqref{361} vanishes.
Then we find easily 
$$(\1+\E)\D=\{(E\zeta,\zeta),\zeta\in\K\},\quad (\1-\E)\D=\{(E^\dg \xi,-\xi),\xi\in\K\}$$
therefore \bes \label{460}
\begin{align} W_k(+\E)(\d_+kk^{-1},0)&=(\d_+kk^{-1},E^{-1}\d_+kk^{-1}),\\ W_k(-\E)(\d_-kk^{-1},0)&=(\d_-kk^{-1},-(E^\dg)^{-1}\d_-kk^{-1}). \end{align}\ees
Inserting \eqref{460} in \eqref{361}, we obtain the desired result \eqref{271}
 \be S=\jp\int d\tau  d\si \l E^{-1}\d_+kk^{-1}, \d_-kk^{-1}\r_\K. \label{491}\ee
 Said in other words, we have proven that every  $\si$-model of the
 type \eqref{491} is indeed of the $\E$-model type  \eqref{361}. This fact permits us to study the integrability and the RG flow of the models \eqref{271} by employing the $\E$-model techniques  described in Sections 2.2 and 2.3. We start with the integrability.

 \ms

We set
\be O(\lm)(\mu,\rho)=\jp F_+(\lm)S^{-1}(\mu+E^\dg \rho)+\jp F_-(\lm)S^{-1}(\mu-E\rho),\label{472}\ee
where $F_\pm(\lm):\K\to\K$ are $\lm$-dependent linear operators which we want to determine.
In particular, this means
$$O(\lm)(E\zeta,\zeta)=F_+(\lm)\zeta, \quad  O(\lm)(E^\dg \xi,-\xi)=F_-(\lm)\xi.$$
The condition of integrability \eqref{381} can then be rewritten as
\be [O(\lm)(E\zeta,\zeta),O(\lm)(E^\dg \xi,-\xi)]_\K=O(\lm)[(E\zeta,\zeta),(E^\dg \xi,-\xi)]_\D\label{ci}\ee
which gives
\begin{multline}[F_+(\lm)\zeta,F_-(\lm)\xi]_\K= \jp F_+(\lm)S^{-1}([E\zeta,E^\dg \xi]_\K+E^\dg[\zeta,\xi]_E)+\\+\jp F_-(\lm)S^{-1}([E\zeta,E^\dg \xi]_\K-E[\zeta,\xi]_E),\label{cib}\end{multline}
where
\be [\zeta,\xi]_E:=[\zeta,E^\dg \xi]_\K-[E\zeta,\xi]_\K.\label{xye}\ee
If the operators $F_\pm(\lm)$ verify the condition \eqref{cib}, then the Lax pair is given by (cf. Section 2.2)
 \bes \label{484}\begin{align} L_+(\lm)&=O(\lm)W_k(+\E)(\d_+kk^{-1},0)=F_+(\lm)E^{-1}\d_+kk^{-1},\\ L_-(\lm)&=O(\lm)W_k(-\E)(\d_-kk^{-1},0)=F_-(\lm)(E^\dg)^{-1}\d_-kk^{-1}.\end{align} \ees
 %%%%%%%%%%%%%%%%%%%%%%%%%%%%%%%%%%%
\subsection{The case $K=SU(2)$}
%%%%%%%%%%%%%%%%%%%%%%%%%%%%%%%%%%%
Now we focus on the  case $K=SU(2)$ and the diagonal operator  $E=D^{-1}$.
We suppose that the operators $F_\pm(\lm)$  are also diagonal in the basis $T_j=-\ri\si_j$, that is
$$F_\pm(\lm)T_j=F^j_\pm(\lm)T_j, \quad j=1,2,3 \quad \textrm{(no summation)}. $$
  The condition of integrability \eqref{cib} then becomes
\bes \label{cond}
\begin{align}D_1D_2F_\pm^1(\lm)F_\mp^2(\lm)&=\jp F_\pm^3(\lm)\l D_3+D_1-D_2\r+\jp F_\mp^3(\lm)\l D_3-D_1+D_2\r\\
D_2D_3F_\pm^2(\lm)F_\mp^3(\lm)&=\jp F_\pm^1(\lm)\l D_1+D_2-D_3\r+\jp F_\mp^1(\lm)\l D_1-D_2+D_3\r\\
D_3D_1F_\pm^3(\lm)F_\mp^1(\lm)&=\jp F_\pm^2(\lm)\l D_2+D_3-D_1\r+\jp F_\mp^2(\lm)\l D_2-D_3+D_1\r.\end{align}\ees
and, if they are fulfilled, the Lax pair \eqref{484}
   reads 
\be L_\pm(\lm)= F_\pm(\lm)D\d_\pm kk^{-1}.\label{elac}\ee

Set
$$ G^j_\pm(\lm):= F^j_\pm(\lm)D_j, \quad K^j:=D_{j+1}+D_{j-1}-D_j, \quad j=1,2,3,$$
where there is no summation over the repeated indices and it is understood that  $3+1=1$ and $1-1=3$. The conditions of integrability \eqref{cond} then become
\bes \label{condb}
\begin{align}2D_3G_\pm^1(\lm)G_\mp^2(\lm)&= K^2G_\pm^3(\lm) +  K^1G_\mp^3(\lm),\\
2D_1G_\pm^2(\lm)G_\mp^3(\lm)&= K^3G_\pm^1(\lm) +  K^2G_\mp^1(\lm),\\
2D_2G_\pm^3(\lm)G_\mp^1(\lm)&= K^1G_\pm^2(\lm) +  K^3G_\mp^2(\lm).\end{align}\ees
  Set furthermore
 \be H^j(\lm)=\frac{G^j_+(\lm)}{G^j_-(\lm)}. \ee
 We  then infer from \eqref{condb}
   \be \label{condd}
  \frac{H^1(\lm)}{H^2(\lm)}=\frac{K^1+K^2H^3(\lm)}{K^2+K^1H^3(\lm)} ,\quad 
\frac{H^2(\lm)}{H^3(\lm)}=\frac{K^2+K^3H^1(\lm)}{K^3+K^2H^1(\lm)}  ,\quad 
\frac{H^3(\lm)}{H^1(\lm)}=\frac{K^3+K^1H^2(\lm)}{K^1+K^3H^2(\lm)} . \ee
The equations \eqref{condd} determine $H^j(\lm)$
 \bes \label{sola}
\begin{align}H^1(\lm)&=\frac{1}{(K^2\lm  +\sqrt{(K^2\lm)^2+1})(K^3\lm  +\sqrt{(K^3\lm)^2+1})}  ,\\
H^2(\lm)&=\frac{1}{(K^3\lm  +\sqrt{(K^3\lm)^2+1})(K^1\lm  +\sqrt{(K^1\lm)^2+1})}  ,\\
H^3(\lm)&=\frac{1}{(K^1\lm  +\sqrt{(K^1\lm)^2+1})(K^2\lm  +\sqrt{(K^2\lm)^2+1})},\end{align}\ees
 
Replacing $G_+^j$ in \eqref{condb} by $H^jG_-^j$, $j=1,2,3$ we obtain
\bes \label{condf}
\begin{align}2D_3H^1(\lm)G_-^1(\lm)G_-^2(\lm)&=K^2 H^3(\lm)G_-^3(\lm) +K^1G_-^3(\lm),\\
2D_1H^2(\lm)G_-^2(\lm)G_-^3(\lm)&=K^3 H^1(\lm)G_-^1(\lm) +K^2 G_-^1(\lm) ,\\
2D_2H^3(\lm)G_-^3(\lm)G_-^1(\lm)&=K^1H^2(\lm)G_-^2(\lm) +K^3 G_-^2(\lm).\\\end{align}\ees
Those equations are easily solved 
giving the final result
\bes \label{condk}
\begin{align}
G_\pm^1(\lm)&= \jp\frac{\sqrt{H^1(\lm)}^{\pm 1}}{\sqrt{D_3D_2}} \sqrt{\l K^1H^2(\lm)+K^3\r\l K^2+\frac{K^1}{H^3(\lm)}\r},\\
G_\pm^2(\lm)&= \jp\frac{\sqrt{H^2(\lm)}^{\pm 1}}{\sqrt{D_1D_3}} \sqrt{\l K^2H^3(\lm)+K ^1\r\l K^3+\frac{K^2}{H^1(\lm)}\r},\\
G_\pm^3(\lm)&= \jp\frac{\sqrt{H^3(\lm)}^{\pm 1}}{\sqrt{D_2D_1}} \sqrt{\l K^3H^1(\lm)+K^2\r\l K^1+\frac{K^3}{H^2(\lm)}\r}.\end{align}\ees

%%%%%%%%%%%%%%%%%%%%%%%%%%%%%%%%
\subsection{ Subcases}
%%%%%%%%%%%%%%%%%%%%%%%%%%%%%%%
We are still in the framework of $K=SU(2)$, but now we consider few  subcases.

\ms 

\no 1) $D_1=D_2=D_3$. In this case we have  also  $K^1=K^2=K^3=D_1$, so that 
$H_1(\lm)=H_2(\lm)=H_3(\lm)$ and  we find
$$G_\pm^1(\lm)=G_\pm^2(\lm)=G_\pm^3(\lm)=\jp\l 1+\frac{\sqrt{D_1^2\lm^2+1}\mp D_1\lm  }{\sqrt{D_1^2\lm^2+1}\pm D_1\lm  }\r.$$
We may change the spectral parameter $\lm$ to a new one $\xi$ according to
$$\xi=-\frac{D_1\lm}{\sqrt{D_1^2\lm^2+1}}$$
which gives 
\be G_\pm^1(\xi)=G_\pm^2(\xi)=G_\pm^3(\xi)=\frac{1}{1\mp \xi}. \label{ts}\ee
Inserting \eqref{ts} into \eqref{elac} reproduces the Zakharov-Mikhailov Lax pair of the principal chiral model.

\ms

\no 

\no 2) $D_1\neq D_2=D_3$. In this case we have $K^2=K^3=D_1$, while $K^1$ is a free parameter (which may take also negative values). We find  $H^2(\lm)=H^3(\lm)$ and 
\begin{multline} G^1_\pm(\lm)=\frac{K^1+K^2H^2(\lm)^{\pm 1}}{2D_2},\\G^2_\pm(\lm)=G^3_\pm(\lm)=\frac{\sqrt{K^2(1+H^2(\lm)^{\pm 2})+2K^1H^2(\lm)^{\pm 1}}}{2\sqrt{D_2}}.\end{multline}
We may choose a new spectral parameter $\xi$ as
$$\xi=\frac{H^2(\lm)-1}{H^2(\lm)+1}.$$
Then 
\be G^1_\pm(\xi)= \frac{D_2\pm (D_1-D_2)\xi}{D_2(1\mp \xi)}, \quad G^{2}_\pm (\xi)=G^{3}_\pm (\xi)=\frac{\sqrt{D_2+(D_1-D_2)\xi^2}}{\sqrt{D_2}(1\mp \xi)}.\label{c2}\ee
For $D_1<D_2=D_3$, we may further change the variables as
$$D_1=\frac{D_2}{\cosh^2{\nu}}, \quad \xi=\frac{\tanh{\nu z}}{\tanh{\nu}}$$
and we thus reproduce  the result of \cite{LW,KS24}
$$G^1_\pm(z)=\frac{\tanh{\nu}}{\tanh{\nu(1\mp z)}}, \quad G^{2}_\pm (z)=G^{3}_\pm (z) =\frac{\sinh{\nu}}{\sinh{\nu(1\mp z)}}.$$

\ms 

\noi 3) $D_1=D_2+D_3$. In this case 
   we have
$K^1=0$, $K^2=2D_3$, $K^3=2D_2$ and 
$H^1(\lm)=H^2(\lm)H^3(\lm)$. This gives 
$$G^1_\pm(\lm)=\sqrt{H^1(\lm)}^{\pm1}, \ G^2_\pm(\lm)=\frac{\sqrt{D_3+D_2H^1(\lm)^{\pm 1}}}{\sqrt{D_1}},  \  
G^3_\pm(\lm)=\frac{\sqrt{D_2+D_3H^1(\lm)^{\pm 1}}}{\sqrt{D_1}}$$
Introducing a new spectral parameter 
$$\xi=\frac{\sqrt{H^1(\lm)}-1}{\sqrt{H^1(\lm)}-1},$$
we finally obtain 
\begin{multline} G^1_\pm(\xi)=\frac{1\pm\xi}{1\mp\xi}, \quad 
G^2_\pm(\xi)= \frac{\sqrt{D_1(1+\xi^2)\pm 2\xi(D_2-D_3)}}{\sqrt{D_1}(1\mp \xi)}, \\
G^3_\pm(\xi)= \frac{\sqrt{D_1(1+\xi^2)\pm 2\xi(D_3-D_2)}}{\sqrt{D_1}(1\mp \xi)}.\label{c3}\end{multline}
 
Note that for $D_1=D_2+D_3$  and, simultaneously, $D_2=D_3$  we are at the intersection of the cases 2) and 3). Indeed, in this very special case we check easily that the formulas   \eqref{c2} and \eqref{c3} coincide and give 
\be G^1_\pm(\xi)=\frac{1\pm\xi}{1\mp\xi}, \quad 
G^2_\pm(\xi)=  
G^3_\pm(\xi)= \frac{\sqrt{ 1+\xi^2 }}{ 1\mp \xi}.\label{c4}\ee
 \ms 

 \noi 4)  $D_1>D_2>D_3$.  We may trade the parameters $D_2,D_3$ for  new parameters $\nu>0,0<k<1$ as in \cite{LW,KS24}
$$ D_2=D_1\frac{\cn^2(\nu;k)}{\dn^2(\nu;k)}, \quad D_3=D_1\cn^2(\nu;k),$$
where  $\sn$, $\cn$ and $\dn$ are the standard Jacobi elliptic functions with the modulus $k$.  At the same time we replace the spectral parameter $\lm$ by another one $z$ according to
$$H_3(\lm)=\frac{\sn(\nu(1+z))}{\sn(\nu(1-z))}.$$
With those changes, the solutions \eqref{condk}
become those of \cite{LW} that is 
\begin{multline} G_\pm^1(z)=\frac{\sn(\nu)\cn(\nu(1\mp z))}{\cn(\nu)\sn(\nu(1\mp z))},\quad G_\pm^2(z)=\frac{\sn(\nu)\dn(\nu(1\mp z))}{\dn(\nu)\sn(\nu(1\mp z))}, \\ G_\pm^3(z)=\frac{\sn(\nu )}{\sn(\nu(1\mp z))}.\end{multline}
 
 %%%%%%%%%%%%%%%%%%%%%%%%
\subsection{RG flow of the non-deformed model}
%%%%%%%%%%%%%%%%%%%%%%%

We still work with the original Cherednik case $K=SU(2)$, $E=D^{-1}$.
We choose the basis of $\D$ as  $T_A$, $A=1,\dots, 6$
$$T_{1,2,3}=(-\ri\si_{1,2,3},0),\quad T_{4,5,6}=(0,-\ri\si_{1,2,3}).$$
 Using the notation of Section 2.3, we then find 
 $$\eta^{AB}=-\jp\bpm 0&0&0&1&0&0\\0&0&0&0&1&0\\0&0&0&0&0&1\\1&0&0&0&0&0\\0&1&0&0&0&0\\
 0&0&1&0&0&0\epm ,\quad \E^{AB}=-\jp\bpm \frac{1}{D_1}&0&0&0&0&0\\0&\frac{1}{D_2}&0&0&0&0\\0&0&\frac{1}{D_3}&0&0&0\\0&0&0&D_1&0&0\\0&0&0&0&D_2&0\\
 0&0&0&0&0&D_3\epm, $$
 $$\M^{AB}=2\bpm \frac{1}{D_2D_3}&0&0&0&0&0\\0&\frac{1}{D_3D_1}&0&0&0&0\\0&0&\frac{1}{D_1D_2}&0&0&0\\0&0&0& \frac{D_2}{D_3}+\frac{D_3}{D_2}-2&0&0\\0&0&0&0&\frac{D_3}{D_1}+\frac{D_1}{D_3} -2&0\\
 0&0&0&0&0&\frac{D_1}{D_2}+\frac{D_2}{D_1}-2\epm. $$
 \begin{multline}
(\E\M\E-\M)^{AB}=\\=\frac{2}{D_1D_2D_3}\bpm -D_1^{-1}K^2K^3&0&0&0&0&0\\0&-D_2^{-1}K^3K^1&0&0&0&0\\0&0&-D_3^{-1}K^1K^2&0&0&0\\0&0&0& D_1K^2K^3&0&0\\0&0&0&0&D_2K^3K^1&0\\
 0&0&0&0&0&D_3K^1K^2\epm. \end{multline}
 Inserting all this in \eqref{fl}, we obtain precisely the RG flows of the couplings $D_1,D_2,D_3$ which were obtained in \cite{LW,KS24,SS14} without using the $\E$-models formalism:
 \be  \frac{dD_1}{d\mu}=-\frac{4K^2K^3}{D_2D_3},\quad 
\frac{dD_2}{d\mu}=-\frac{4K^3K^1}{D_3D_1},\quad \frac{dD_3}{d\mu}=-\frac{4K^1K^2}{D_1D_2}.
\ee

% Consider a line vector
% $$v_\pm =\bpm u_+^2u_+^3 F_\pm^1 & u_-^1u_-^3 F_\pm^2&-u_-^1u_+^2 F_\pm^3\epm.$$
% We find $v_\pm M_\pm=0$ which means (cf. \eqref{inf})
 %$$v_\pm F_\pm=u_+^2u_+^3(F_\pm^1)^2+u_-^1u_-^3(F_\pm^2)^2-u_-^1u_+^2 (F_\pm^3)^2=0.$$

  %%%%%%%%%%%%%%%%%%%%%%%%
\section{Poisson-Lie deformed Cherednik model}
\subsection{General case}
%%%%%%%%%%%%%%%%%%%%%%%

We now determine  the Drinfeld double, the involution $\E:\D\to\D$ and the isotropic subalgebra $\B$
in such a way that the Poisson-Lie  deformed $\si$-model \eqref{290b} can be written
in the $\E$-model form \eqref{361}.  

\ms

The  Drinfeld double is the Lu-Weinstein one \cite{LuW}, that is it has the structure of the complexified group  $K^\bc$ viewed as the real group. If we parametrize the elements of $\D=\K^\bc$ as $u+\ri v$, $u,v\in\K$,
then the   invariant bilinear form $(.,.)_\D$ is defined as 
\be (x+\ri y,u+\ri v)_\D:=\frac{1}{\nu} (x,v)_\K+\frac{1}{\nu}(y,u)_\K. \label{959}\ee
Here $\nu$ is  a real positive parameter.

\ms 

 Let $E:\K\to\K$ be the invertible  linear operator with the invertible symmetric part $S$ and define  the linear selfadjoint involution $\E_{\nu}:\K^\bc\to \K^\bc$ by 
    \be\E_{\nu}(u+\ri v)= -(u+\ri v)+(E+\ri\nu) S^{-1}\l \nu^{-1}E^\dg v+u\r. \label{961}\ee

    Finally, we take  $B=AN$ for the  half-dimensional isotropic subgroup  $B$ 
of $D$. Recall that $A\subset K^\bc$ is the Abelian subgroup such that 
Lie($A$)$=\ri\T$, where $\T$ is the Cartan subalgebra of $\K$. As far as $N$ is concerned, it is the nilpotent subgroup of $K^\bc$, obtained by exponentiation of the positive step generators $E^\al$ of $\K^\bc$. For example, for the case of the matrix groups $K=SU(N)$ and $K^\bc=SL(N,\bc)$, the group
$AN$ consists of complex upper-triangular matrices with real positive numbers 
on the diagonal the product of which is equal to $1$.
 We note that contrary to the previous section, the subgroup $B$ is not Abelian.   

\medskip 

It is convenient to parametrize the elements of the Lie algebra $\B$ in terms of the elements of $\K$. This is done by using certain linear operator $R:\K^\bc\to\K^\bc$  called the Yang-Baxter operator. In terms of the standard Chevalley basis of $\K^\bc$, it is defined by 
$$RE^\al=-{\rm sign}(\al)\ri E^\al,\quad RH_j=0.$$
It turns out that the restriction of $R$ on $\K$ gives a linear operator $R:\K\to\K$. Moreover, every element $\mu\in\B$ can be uniquely represented by an element $x\in\K$ via the relation
$$\mu =(R-\ri)x.$$
This fact is useful if we wish to determine the expressions $W_k(\pm \E)\d_\pm kk^{-1}$ for $k\in K$. To do that we first find  
$$(\1+\E_\nu)\D=(E+\ri\nu)\K,\quad (\1-\E_\nu)\D=(E^\dg-\ri \nu)\K,$$ 
in particular, we have \bes \label{665}\begin{align}
\jp(\1+\E_E)(u+\ri v)&=\jp(E+\ri\nu) S^{-1}\l u+ \nu^{-1}E^\dg v\r,\\
\jp(\1-\E_E)(u+\ri v)&=
\jp(E^\dg-\ri\nu) S^{-1}\l u- \nu^{-1}E v\r.\end{align}\ees

Then  we determine $u,v\in\K$ such that it holds
$$\d_+kk^{-1}=(E+\ri\nu)u+Ad_k(R-\ri) v.$$
We find 
$$v=\l \nu^{-1}E\Ad_k +\Ad_k R\r^{-1}\d_+kk^{-1},\quad u=\nu^{-1}\Ad_k v, $$
therefore 
\be W_k(+\E)\d_+ kk^{-1}= (E+\ri\nu)(E+\nu R_{k^{-1}})^{-1}\d_+kk^{-1}, \quad R_{k^{-1}}=\Ad_k R\Ad_{k^{-1}}.\label{q+}\ee
Similarly we find 
\be W_k(-\E)\d_- kk^{-1}=(E^\dg -\ri\nu)(E^\dg-\nu R_{k^{-1}})^{-1}\d_-kk^{-1}.\label{q-}\ee
Now we are ready to make explicit the action \eqref{361} for the Drinfeld double $K^\bc$,
the operator $\E_\nu$ and the subgroup $B=AN$.
We fix the gauge symmetry $l\to lb$ by setting $l=k$, $k\in K$, which
is possible thanks to the validity of the Iwasawa decomposition $K^\bc=KAN$. As the consequence of the gauge fixing,  the WZ term in \eqref{361} vanishes.
Inserting  \eqref{q+}, \eqref{q-} in \eqref{361} and  taking into account also the antisymmetry of $R$ with respect to the Killing-Cartan form $(.,.)_\K$, we obtain the desired result \eqref{290b}
  \be S_\nu(k)=\jp\int d\tau  d\si \l (E+\nu R_{k^{-1}})^{-1}\d_+kk^{-1}, \d_-kk^{-1}\r_\K. \label{710}\ee
 Said in other words, we have proven that the Poisson-Lie deformed $\si$-model
 \eqref{710} is indeed of the $\E$-model type  \eqref{361}. This fact permits us to study the integrability and the RG flow of the  model  \eqref{710} by employing the $\E$-model techniques  described in Sections 2.2 and 2.3. We start with the integrability.

  \ms

Now  we look for the family of linear operators $O(\lm)$ needed to establish the integrability of the deformed  model \eqref{710}.  Taking inspiration from \eqref{665}, we start with the ansatz (cf. also \eqref{472})
$$O(\lm)(u+\ri v)=\jp F_+(\lm)S^{-1}(u+\nu^{-1}E^\dg v)+
\jp F_-(\lm)S^{-1}(u-\nu^{-1}Ev),$$
where $F_\pm(\lm):\K\to\K$ are $\lm$-dependent linear operators.
In particular, we have 
$$O(\lm)(E+\ri\nu) x=F_+(\lm)x, \quad O(\lm)(E^\dg -\ri\nu) y=F_-(\lm)y, \quad x,y\in\K.$$
 The condition of integrability \eqref{381} then  becomes  
$$[O(\lm)(E+\ri\nu) x,O(\lm)(E^\dg-\ri\nu) y]_\K=O(\lm)[(E+\ri\nu) x,(E^\dg-\ri\nu) y]_{\K^\bc}$$
which gives
\begin{multline}[F_+(\lm)x,F_-(\lm)y]= \jp F_+(\lm)S^{-1}([Ex,E^\dg y]_\K+\nu^2[x,y]_\K+E^\dg[x,y]_E)+\\+\jp F_-(\lm)S^{-1}([Ex,E^\dg y]_\K+\nu^2[x,y]_\K-E[x,y]_E),\label{cipl}\end{multline}
where the bracket $[x,y]_E$ was introduced in \eqref{xye}. We observe that the conditions \eqref{cipl} are $\nu$-deformed versions of the conditions \eqref{cib}.

\ms 

If the operators $F_\pm(\lm)$ verify the integrability  condition \eqref{cipl}, then the Lax pair is given by (cf. Section 2.2)
 \bes \label{736}\begin{align} L_+(\lm)&=O(\lm)W_k(+\E)\d_+kk^{-1}=F_+(\lm)(E+\nu R_{k^{-1}})^{-1}\d_+kk^{-1},\\ L_-(\lm)&=O(\lm)W_k(-\E)(\d_-kk^{-1},0)=F_-(\lm)(E^\dg-\nu R_{k^{-1}})^{-1}\d_-kk^{-1}.\end{align} \ees

%%%%%%%%%%%%%%%%%%%%%%%
\subsection{Case $K=SU(2)$}
%%%%%%%%%%%%%%%%%%%%%%%%

Consider again the case  $SU(2)$ and $E=D^{-1}$ is diagonal in the  basis of $su(2)$ given by the  matrices $-\ri\si_1,-\ri\si_2,-\ri\si_3$. We suppose that $F_\pm(\lm)$ are also diagonal, the deformed condition of integrability \eqref{cipl} then becomes
 \begin{multline}\ \\D_1D_2F_\pm^1(\lm)F_\mp^2(\lm)=\\=\jp F_\pm^3(\lm)\l D_3+\nu^2D_1D_2D_3+D_1-D_2\r+\jp F_\mp^3(\lm)\l D_3+\nu^2D_1D_2D_3-D_1+D_2\r,\\D_2D_3F_\pm^2(\lm)F_\mp^3(\lm)=\\=\jp F_\pm^1(\lm)\l D_1+\nu^2D_1D_2D_3+D_2-D_3\r+\jp F_\mp^1(\lm)\l D_1+\nu^2D_1D_2D_3-D_2+D_3\r,\\D_3D_1F_\pm^3(\lm)F_\mp^1(\lm)=\\=\jp F_\pm^2(\lm)\l D_2+\nu^2D_1D_2D_3+D_3-D_1\r+\jp F_\mp^2(\lm)\l D_2+\nu^2D_1D_2D_3-D_3+D_1\r.\end{multline} 
and the Lax pair \eqref{710} becomes
\be L_\pm(\lm)= F_\pm(\lm)(1\pm\nu DR_{k^{-1}})^{-1}D\d_\pm kk^{-1}.\label{elad}\ee

\ms 

Similarly as in the non-deformed case, set  
$$ G^j_\pm(\lm):= F^j_\pm(\lm)D_j, \quad K^j_\nu:=D_{j+1}+D_{j-1}+\nu^2D_1D_2D_3-D_j, \quad j=1,2,3$$
where there is no summation over the repeated indices and it is understood that  $3+1=1$ and $1-1=3$. The $\nu$-deformed conditions of integrability  then become almost identical as the non-deformed ones \eqref{condb}
\bes \label{condt}
\begin{align}2D_3G_\pm^1(\lm)G_\mp^2(\lm)&= K^2_\nu G_\pm^3(\lm) +  K^1_\nu G_\mp^3(\lm),\\
2D_1G_\pm^2(\lm)G_\mp^3(\lm)&= K^3_\nu G_\pm^1(\lm) +  K^2_\nu G_\mp^1(\lm),\\
2D_2G_\pm^3(\lm)G_\mp^1(\lm)&= K^1_\nu G_\pm^2(\lm) +  K^3_\nu G_\mp^2(\lm).\end{align}\ees
Indeed, the whole effect of the deformation consists only in modifying  the values of the coefficients $K^j$, therefore the deformed conditions are solved precisely in the same way as in the non-deformed case
 \bes \label{condl}
\begin{align}
G_\pm^1(\lm)&= \jp\frac{\sqrt{H^1_\nu(\lm)}^{\pm 1}}{\sqrt{D_2D_1}} \sqrt{\l K^1_\nu H^2_\nu(\lm)+K_{\nu}^3\r\l K^2_\nu +\frac{K^1_\nu}{H_\nu^3(\lm)}\r},\\
G_\pm^2(\lm)&= \jp\frac{\sqrt{H^2_\nu(\lm)}^{\pm 1}}{\sqrt{D_2D_1}} \sqrt{\l K^2_\nu H^3_\nu(\lm)+K_{\nu}^1\r\l K^3_\nu +\frac{K^2_\nu}{H_\nu^1(\lm)}\r},\\
G_\pm^3(\lm)&= \jp\frac{\sqrt{H^3_\nu(\lm)}^{\pm 1}}{\sqrt{D_2D_1}} \sqrt{\l K^3_\nu H^1_\nu(\lm)+K_{\nu}^2\r\l K^1_\nu +\frac{K^3_\nu}{H_\nu^2(\lm)}\r},\end{align}\ees
where 
 \bes \label{solb}
\begin{align}H^1_\nu (\lm)&=\frac{1}{(K^2_\nu \lm  +\sqrt{(K^2_\nu\lm)^2+1})(K^3_\nu \lm  +\sqrt{(K^3_\nu\lm)^2+1})}  ,\\
H^2_\nu (\lm)&=\frac{1}{(K^3_\nu \lm  +\sqrt{(K^3_\nu\lm)^2+1})(K^1_\nu \lm  +\sqrt{(K^1_\nu\lm)^2+1})}  ,\\
H^3_\nu (\lm)&=\frac{1}{(K^1_\nu \lm  +\sqrt{(K^1_\nu\lm)^2+1})(K^2_\nu \lm  +\sqrt{(K^2_\nu\lm)^2+1})}.\end{align}\ees
 
\ms

\subsection{The deformed RG flow}
We still work with the original Cherednik case $K=SU(2)$, $E=D^{-1}$.
We choose the basis of $\D=\K^\bc$ as  $T_A$, $A=1,\dots, 6$
$$T_{1,2,3}=-\ri\si_{1,2,3},\quad T_{4,5,6}=\si_{1,2,3}$$
and we determine the quantities defined in Section 2.3
 $$\eta^{AB}=-\jp\bpm 0&0&0&1&0&0\\0&0&0&0&1&0\\0&0&0&0&0&1\\1&0&0&0&0&0\\0&1&0&0&0&0\\
 0&0&1&0&0&0\epm ,\quad \E^{AB}=-\jp\bpm \frac{1}{D_1}&0&0&0&0&0\\0&\frac{1}{D_2}&0&0&0&0\\0&0&\frac{1}{D_3}&0&0&0\\0&0&0&D_1&0&0\\0&0&0&0&D_2&0\\
 0&0&0&0&0&D_3\epm, $$
  $$[[\1,\1]]^{AB}=\bpm  -4\nu^2&0&0&0&0&0\\0& -4\nu^2&0&0&0&0\\0&0& -4\nu^2&0&0&0\\0&0&0&4&0&0\\0&0&0&0& 4&0\\
 0&0&0&0&0&4\epm $$
\begin{multline}[[\E,\E]]^{AB}=\\= 2\bpm \frac{1}{D_2D_3}+\nu^4D_2D_3&0&0&0&0&0\\0&\frac{1}{D_3D_1}+\nu^4D_3D_1&0&0&0&0\\0&0&\frac{1}{D_1D_2}+\nu^4D_1D_2&0&0&0\\0&0&0&  \frac{D_2}{D_3}+\frac{D_3}{D_2} &0&0\\0&0&0&0&\frac{D_3}{D_1}+\frac{D_1}{D_3}&0\\
 0&0&0&0&0&\frac{D_1}{D_2}+\frac{D_2}{D_1} \epm.\end{multline}
 
  \begin{multline}
(\E\M\E-\M)^{AB}=\\=\frac{2}{D_1D_2D_3}\bpm -D_1^{-1}K_\nu^2K_\nu ^3&0&0&0&0&0\\0&-D_2^{-1}K_\nu^3K_\nu^1&0&0&0&0\\0&0&-D_3^{-1}K_\nu^1K_\nu^2&0&0&0\\0&0&0& D_1K_\nu^2K_\nu^3&0&0\\0&0&0&0&D_2K_\nu^3K_\nu^1&0\\
 0&0&0&0&0&D_3K_\nu^1K_\nu^2\epm. \end{multline}

 Inserting all this in \eqref{fl}, we recover the deformed  RG flows, which were obtained in \cite{KS24} without using the $\E$-models formalism (our $D_j$'s are $\tilde I_j$'s of \cite{KS24}):
 \be  \frac{dD_1}{d\mu}=-\frac{4K_\nu^2K_\nu^3}{D_2D_3},\quad 
\frac{dD_2}{d\mu}=-\frac{4K_\nu^3K_\nu^1}{D_3D_1},\quad \frac{dD_3}{d\mu}=-\frac{4K_\nu^1K_\nu^2}{D_1D_2}.
\ee

  %%%%%%%%%%%%%%%%%%%%%%%%%%%%%%%%%%%%%%%%%%%%
  \section{Poisson-Lie-WZ  deformed  Cherednik model}
  \subsection{General case}
%%%%%%%%%%%%%%%%%%%%%%%%%%%%%%%%%%%%%%%%%%%%%%

We now determine  the Drinfeld double, the involution $\E:\D\to\D$ and the isotropic subalgebra $\B$
in such a way that the Poisson-Lie-WZ deformed $\si$-model \eqref{293} can be written
in the $\E$-model form \eqref{361}.  

\ms

The  Drinfeld double is  again the complexified group  $K^\bc$ viewed as the real group, so that we can parametrize the elements of $\D=\K^\bc$ as $u+\ekn v$, $u,v\in\K$. But now the 
 invariant bilinear form $(.,.)_\D$ is not the Lu-Weinstein one  but it is given by 
 \be  (x+\ekn y,u+\ekn v)_\D:= \ka(x,u)_\K-\ka (y,v)_\K,\quad  x,y,u,v\in \K.\label{1153}\ee
%\begin{multline} (x+\ri y,u+\ri v)_\D:=\ka\cot{(\ka\nu)} ((x,v)_\K+ (y,u)_\K)+\ka(x,u)_\K-\ka (y,v)_\K,\\ \  x,y,u,v\in \K.\label{1153}\end{multline}
 Let $E:\K\to\K$ be the invertible  linear operator. We define  a linear self-adjoint involution $\E_{\ka,\nu}:\K^\bc\to \K^\bc$ by 
 \bes\label{1154}\begin{align}\E_{\ka,\nu}u &=\l e^{\ka E}+e^{-\ka E^\dg}-2e^{-\ri\ka\nu}\r \l \epk-\emkd\r^{-1}u, \quad u\in\K,\\
  \E_{\ka,\nu}\ekn v &=\l e^{-\ka E}+e^{\ka E^\dg}-2e^{\ri\ka\nu}\r \l \emk-\epkd\r^{-1}\ekn v, \quad v\in\K.\end{align}\ees 
  It can be checked  that in the limit $\ka\to 0$, the  formulas \eqref{1153} and \eqref{1154} boil down to the formulas \eqref{959} and \eqref{961}.

\ms

  In order to work out  the $\si$-model action \eqref{361} in this particular case, we need 
  the half-dimensional subgroup $B$ of the Drinfeld double $K^\bc$, which would 
  be isotropic with respect to the bilinear form \eqref{1153} and which would boil down to the upper-triangular subgroup $AN$ in the limit $\ka\to 0$.
  Such subgroup $B$ was described in \cite{K20}, we start with the description of its Lie algebra $\B\subset \K^\bc$ by setting
\be \B=\left\{ \frac{e^{-\ri \ka\nu}-e^{-\ka\nu R}}{\sin{\ka\nu}}y, \ y\in\K\right\},\label{1176}  \ee
where $R$ is the Yang-Baxter operator.

\ms

The isotropic  linear subspace $\B\subset \K^\bc$ is really the Lie subalgebra because
of the identity
 \be \left[\frac{e^{-\ri \ka\nu}-e^{-\ka\nu R}}{\sin{\ka\nu}}x,
\frac{e^{-\ri \ka\nu}-e^{-\ka\nu R}}{\sin{\ka\nu}}y\right]=
\frac{e^{-\ri \ka\nu}-e^{-\ka\nu R}}{\sin{\ka\nu}}[x,y]_{R,\ka,\nu},\quad x,y\in \K.\label{1186}\ee
where $[.,.]_{R,\ka,\nu}$ is an alternative Lie bracket on the vector space $\K$  defined in terms of the standard Lie bracket $[.,.]_\K$  as
\be [x,y]_{R,\ka,\nu}:=\left[\frac{\cos{(\ka\nu)}-e^{-\ka\nu R}}{\sin{\ka\nu}}x,y\right]_\K+\left[x,\frac{\cos{(\ka\nu)}-e^{-\ka\nu R}}{\sin{\ka\nu}}y\right]_\K.\label{Rrhobracket}\ee
Note  that 
\be \lim_{\ka\to 0}[x,y]_{R,\ka,\nu}=[x,y]_R:=[Rx,y]_\K+[x,Ry]_\K. \label{1190}\ee
and the identity \eqref{1186} becomes in this limit the standard Yang-Baxter identity
\be [(R-\ri)x,(R-\ri)y]_{\K^\bc}=(R-\ri)[x,y]_R.\label{1191}\ee
  For our purposes, we don't need to know   details about  the structure of the  corresponding isotropic subgroup $B$, we just  note for completeness that $B$ is a  semi-direct product of a certain  real form of the complex Cartan torus  with the upper-triangular nilpotent subgroup $N\subset K^\bc$.
Moreover, it  turns out that the space of cosets $D/B$ can be identified with the group $K$, therefore we can fix the gauge 
  in the second order action \eqref{361} by replacing the $K^\bc$-valued field $l$ by the   $K$-valued one $k$. However, contrary to the cases
  studied in the two previous sections, this replacing does not make to vanish the WZ term in \eqref{361}.

\medskip 
 
It remains to determine the expressions $W_k(\pm \E)\d_\pm kk^{-1}$ for $k\in K$. To do that, we first find  
\bes\label{822}\begin{align}
 \jp(\1+\E_{\ka,\nu })(u+\ekn v)&=(\1-e^{-\ka (E+\ri\nu)})(1-\emkd\emk)^{-1}(u+\emkd v) \\\jp(\1-\E_{\ka,\nu })(u+\ekn v)&=(\1-e^{\ka (E^\dg-\ri\nu)})(1-\epk\epkd)^{-1}(u+\epk v)
\end{align}\ees
 This means that the   projection operators 
  $\jp(\1\pm\E_{\ka,\nu})$ project, respectively, on the  half-dimensional subspaces 
   $(\1-e^{-\ka (E+\ri\nu)})\K$ and $(\1-e^{\ka (E^\dg-\ri\nu)})\K$.  \quad  

 \ms 

To determine e.g. $W_k(+\E)$, we have to find  $u,v\in\K$ such that it holds
$$\d_+kk^{-1}=(\1-e^{-\ka (E+\ri\nu)})u+Ad_k\frac{e^{-\ri \ka\nu}-e^{-\ka\nu R}}{\sin{\ka\nu}} v.$$
We find 
$$v=\sin{(\ka\nu)}\Ad_{k^{-1}}\l \epk -e^{-\ka\nu R_{k^{-1}}}\r^{-1}\d_+kk^{-1},\quad u=\frac{e^{\ka E}}{\sin{(\ka\nu)}}\Ad_k v, $$
therefore 
\be \label{qw+}  W_k(+\E)\d_+ kk^{-1}= (\1-e^{-\ka (E+\ri\nu)}) \l 1-e^{-\ka\nu R_{k^{-1}}}\emk\r^{-1}\d_+ kk^{-1}.
\ee
Similarly we find 
\be  W_k(-\E)\d_- kk^{-1}= (\1-e^{\ka (E^\dg-\ri\nu)}) \l 1-e^{-\ka\nu R_{k^{-1}}}\epkd\r^{-1}\d_- kk^{-1}. \label{qw-}\ee
Inserting  \eqref{qw+}, \eqref{qw-} in \eqref{361}  
 and taking into account also the antisymmetry of $R$,  we obtain the desired result \eqref{293}
      \begin{multline} S_{\ka,\nu}(k)=\frac{\ka}{4}\int d\tau d\si  \biggl( \frac{1+e^{-\ka\nu R_{k^{-1}}}e^{-\ka E }}{1-e^{-\ka\nu R_{k^{-1}}}e^{-\ka E }}\partial_+ kk^{-1},\partial_- k k^{-1}\biggr)_\K +\\+\frac{\ka}{4}\int d^{-1}\oint (dkk^{-1},[\partial_\sigma kk^{-1},dkk^{-1}])_\K.\label{1205}\end{multline}
 Said in other words, we have proven that the Poisson-Lie-WZ deformed $\si$-model
 \eqref{1205} is indeed of the $\E$-model type  \eqref{361}. This fact permits us to study the integrability  of the  model  \eqref{1205} by employing the $\E$-model techniques  described in Sections 2.2 and 2.3.  
 
  \ms 
   
We look for the family of linear operators $O(\lm)$ needed to establish the integrability of the deformed  model \eqref{1205}.  Taking inspiration from \eqref{822}, we start with an ansatz 
\begin{multline} O(\lm)(u+\ekn v)=\\=  \ka F_+(\lm) (1-\emkd\emk)^{-1}(u+\emkd v) -
\ka  F_-(\lm) (1-\epk\epkd)^{-1}(u+\epk v),\end{multline}
 where $F_\pm(\lm):\K\to\K$ are $\lm$-dependent linear operators.
In particular, we have 
$$O(\lm)(\1-e^{-\ka (E+\ri\nu)}) x=\ka F_+(\lm) x, \quad O(\lm)(\1-e^{\ka (E^\dg-\ri\nu)}) y=- \ka F_-(\lm) y, \quad x,y\in\K.$$
 The condition of integrability \eqref{381} then  becomes  
\begin{multline}
    [O(\lm)(\1-e^{-\ka (E+\ri\nu)}) x,O(\lm)(\1-e^{\ka (E^\dg-\ri\nu)}) y]_\K=\\=O(\lm)[(\1-e^{-\ka (E+\ri\nu)}) x,(\1-e^{\ka (E^\dg-\ri\nu)}) y]_{\K^\bc},
\end{multline}
which gives
\begin{multline} [F_+(\lm)x,F_-(\lm)y]=\\=\ka^{-1}( F_+(\lm)+ F_-(\lm))(\emk-\epkd)^{-1}(2\cos{(\ka\nu)}[\emk x,\epkd y]-[\emk x,y]-[x,\epkd y])+\\+\ka^{-1}( F_+(\lm)\epk\epkd+F_-(\lm))(1-\epk\epkd)^{-1}([x,y]-[\emk x,\epkd y]).\label{ciplw}\end{multline}
Here  all commutators are of the type $[.,.]_\K$.

\ms 

If the operators $F_\pm(\lm)$ verify the integrability  condition \eqref{ciplw}, then the Lax pair is given by (cf. Section 2.2)
 \bes \label{866}\begin{align} L_+(\lm)&=O(\lm)W_k(+\E)\d_+kk^{-1}=F_+(\lm) \l 1-e^{-\ka\nu R_{k^{-1}}}\emk\r^{-1}\d_+ kk^{-1},\\ L_-(\lm)&=O(\lm)W_k(-\E)\d_-kk^{-1}=-F_-(\lm)\l 1-e^{-\ka\nu R_{k^{-1}}}\epkd\r^{-1}\d_- kk^{-1}.\end{align} \ees

%%%%%%%%%%%%%%%%%%%%%%%
\subsection{Case $K=SU(2)$}
%%%%%%%%%%%%%%%%%%%%%%%%

Consider again the case  $SU(2)$ and $E=D^{-1}$ is diagonal in the  basis of $su(2)$ given by the  matrices $-\ri\si_1,-\ri\si_2,-\ri\si_3$. We suppose that $F_\pm(\lm)$ are also diagonal, the doubly deformed condition of integrability \eqref{ciplw} then becomes
 \be \label{876bis}
 \ka\sinh{\l\ka D_{j-1}^{-1}\r} F_\pm^j(\lm)F_\mp^{j+1}(\lm)=\jp c_\pm^{j+1}l_j^{\mp 1}F_\pm^{j-1}(\lm)  + \jp c_\mp^jl_{j+1}^{\pm 1}F_\mp^{j-1}(\lm), \quad  j=1,2,3,\ee
 where 
 \be c_\pm ^j=l_j^{\pm 1}(l_{j+1}^{\pm 1}-2\cos{(\ka\nu)})+1 +l_{j-1}^{\pm 1}(l_j^{\pm 1}-l_{j+1}^{\pm 1}), \quad l_j=e^{\frac{\ka}{D_j}}.\label{883}\ee
 Set
\be  H^j(\lm)=\frac{F_+^j(\lm)}{F^j_-(\lm)}, \quad j=1,2,3.\ee
The equations  \eqref{876bis} then imply
\be  \frac{H^j(\lm)}{H^{j+1}(\lm)}=  \frac{c_+^{j+1} l_j^{-1}H^{j-1}(\lm) +l_{j+1}c_-^j}{l_jc_-^{j+1}+c_+^1l_{j+1}^{-1}H^{j-1}(\lm)}, \quad j=1,2,3.\label{898}\ee
 \begin{multline} \label{901}
\ka\sinh{\l\ka D_{j-1}^{-1}\r} H^j(\lm)F_-^j(\lm)F_-^{j+1}(\lm)=\\=\jp c_+^{j+1}l_j^{- 1}H^{j-1}(\lm)F_-^{j-1}(\lm)  + \jp c_-^jl_{j+1}F_-^{j-1}(\lm), \quad  j=1,2,3.\end{multline}
%{\color{blue}\bes\begin{align}\ka\sinh{\l\ka D_3^{-1}\r}H^1(\lm)F_-^1(\lm)F_-^2(\lm)&= \jp c_+^{2}l_1^{- 1}H^3(\lm)F_-^3(\lm) +\jp c_-^1l_{2}F_-^3(\lm),\\
%\ka\sinh{\l\ka D_1^{-1}\r}H^2(\lm)F_-^2(\lm)F_-^3(\lm)&=\jp c_+^{3}l_2^{- 1} H^1(\lm)F_-^1(\lm) +\jp c_-^2l_{3}F_-^1(\lm) ,\\
%\ka\sinh{\l\ka D_2^{-1}\r}H^3(\lm)F_-^3(\lm)F_-^1(\lm)&=\jp c_+^{1}l_3^{-1}H^2(\lm)F_-^2(\lm) +\jp c_-^3l_{1} F_-^2(\lm).\end{align}\ees }
From \eqref{901} we solve $F_-^j(\lm)$ in terms of $H^j(\lm)$ as 
%{\color{blue}\be  F_-^2(\lm)=\frac{1}{2\ka}\sqrt{\frac{(c_-^1l_2+c_+^2l_1^{-1}H^3(\lm))(c_-^2l_3+c_+^3l_2^{-1}H^1(\lm))}{\sinh{\l\ka D_3^{-1}\r}\sinh{\l\ka D_1^{-1}\r}H_1(\lm)H_2(\lm)}}\ee}
\be \label{907}F_-^j(\lm)=\frac{1}{2\ka}\sqrt{\frac{(c_-^{j-1}l_j+c_+^jl_{j-1}^{-1}H^{j+1}(\lm))(c_-^jl_{j+1}+c_+^{j+1}l_j^{-1}H^{j-1}(\lm))}{\sinh{\l\ka D_{j+1}^{-1}\r}\sinh{\l\ka D_{j-1}^{-1}\r}H_{j-1}(\lm)H_j(\lm)}}.\ee
It remains to solve \eqref{898}.
 Without a loss of generality, we may look for a solution of  the form
$$H_1(\lm)=l_2l_3g_2(\lm)g_3(\lm), \quad H_2(\lm)=l_3l_1g_3(\lm)g_1(\lm), \quad H_3(\lm)=l_1l_2g_1(\lm)g_2(\lm),$$
which transforms \eqref{898} into 
\be \e_{jkl}c_+^kg_l(\lm)=\e_{jkl}c_-^kg_l(\lm)^{-1}, \quad j=1,2,3.\label{ei}\ee
Here $\e_{jkl}$ is the standard Levi-Civitta symbol.
Introducing  vectors $\bs c_\pm$, $\bs g(\lm)$, $\bs g(\lm)^{-1}$ as
$$\bs c_\pm=\bpm c_\pm^1\\c_\pm^2\\c_\pm^3\epm, \quad \bs g(\lm)^{\pm 1}=\bpm g_1(\lm)^{\pm 1}\\g_2(\lm)^{\pm 1}\\g_3(\lm)^{\pm 1}\epm,$$
we may rewrite the basic integrability  condition \eqref{ei} in terms of the cross products
\be \bs c_+ \times \bs g(\lm)=\bs c_-\times \bs g(\lm)^{-1}.\label{cro}\ee
Our strategy to solve \eqref{cro} will be as follows: we first give one particular solution and then we show that it is the only one.

\ms 
 
Let at least  two of the parameters $D_1,D_2,D_3$  be equal to each other, without a loss of generality we may suppose $D_2=D_3$,  
that is $l_2=l_3=l$. In this case we have\footnote{Note that $c_\pm$ and $c^1_-$ are positive numbers whatever is $l$ and $l_1$, however $c^1_+$ may be positive, negative or null. However if, $l\leq 1+\sqrt{2 -2\cos{(\ka\nu)}}$, then $c^1_+$ is also always positive.} 
$$c_\pm^1=1-l^{\pm 2}+2l_1^{\pm 1}l^{\pm 1}-2l_1^{\pm}\cos{(\ka\nu)}, \quad c_\pm^2=c_\pm^3=c_\pm:=l^{\pm 2}-2l^{\pm 1}\cos{(\ka\nu)}+1$$
and the basic  condition  \eqref{ei} becomes
\bes \label{919}\begin{align} c_+g_3(\lm)-c_+g_2(\lm)&=c_-g_3(\lm)^{-1}-c_-g_2(\lm)^{-1} \\c_+g_1(\lm)-c_+^1g_3(\lm)&=c_-g_1(\lm)^{-1}-c_-^1g_3(\lm)^{-1}\\c_+^1g_2(\lm)-c_+g_1(\lm)&=c_-^1g_2(\lm)^{-1}-c_-g_1(\lm)^{-1}.\end{align}\ees
We may solve the first one simply by setting  $g_2(\lm)=g_3(\lm)=g(\lm)$, in which case the 
second and the third one fuse into the same expression
\be c_+g_1(\lm)-\frac{c_-}{g_1(\lm)}=c_+^1g(\lm)-\frac{c_-^1}{g(\lm)}.\label{922}\ee
The condition \eqref{922} is then solved by 
\be g_1(\lm)=\frac{\lm -\sqrt{\lm^2+4c_+c_-}}{2c_+}, \quad g(\lm)=\frac{2c_-^1}{\sqrt{\lm^2+4c_+^1c_-^1}-\lm},\label{924}\ee

\ms 

Now we show that the solution just described is the only one. Said differently,
we show that if all $D_j$ are different, that is $D_1\neq D_2\neq D_3\neq D_1$, then the
basic integrability condition \eqref{ei} does not have any solution. To prove that, we note that if $\bs g(\lm)$ is to solve \eqref{cro},  it   must exist a scalar $A(\lm)$ such that
\be \bs g(\lm)\times \bs g(\lm)^{-1}=A(\lm) \bs c_+\times \bs c_-.\label{al}\ee
Indeed, the relation  \eqref{cro} implies 
$$\bs g(\lm)^{-1}.(\bs c_+\times \bs g(\lm))=0=\bs g(\lm).(\bs c_-\times \bs g(\lm)^{-1}),$$
or, equivalently,
$$\bs c_+.( \bs g(\lm)\times \bs g(\lm)^{-1})=0=\bs c_-.( \bs g(\lm)\times \bs g(\lm)^{-1}),$$
therefore  \eqref{al} easily follows. Written in components, \eqref{al} becomes
\be \frac{g_2(\lm)}{g_3(\lm)}- \frac{g_3(\lm)}{g_2(\lm)} =A(\lm) q_1,\ee
\be \frac{g_3(\lm)}{g_1(\lm)}- \frac{g_1(\lm)}{g_3(\lm)}=A(\lm) q_2, \ee
\be \frac{g_1(\lm)}{g_2(\lm)}- \frac{g_2(\lm)}{g_1(\lm)}=A(\lm) q_3, \ee
where
$$q_j=\e_{jkl}c_{+k}c_{-l}, \quad j=1,2,3.$$
It follows
\be \frac{g_{j+1}(\lm)}{g_{j-1}(\lm)}=\jp A(\lm)q_j\pm \sqrt{\js A(\lm)^2q_j^2+1}, \quad j=1,2,3.\ee
Then we have
\begin{multline} \frac{g_2(\lm)}{g_3(\lm)} \frac{g_3(\lm)}{g_1(\lm)}\frac{g_1(\lm)}{g_2(\lm)}=1=\l \jp A(\lm)q_1\pm \sqrt{\js A(\lm)^2q_1^2+1}\r \times \\\times \l \jp A(\lm)q_2\pm \sqrt{\js A(\lm)^2q_2^2+1}\r  \l \jp A(\lm)q_3\pm \sqrt{\js A(\lm)^2q_3^2+1}\r.\label{945}\end{multline}
Whatever is the combinations of signs on the right-hand-side, we arrive at the conclusion that the second equality in \eqref{945} may hold only if one of the numbers $q_1,q_2,q_3$ vanishes. Without loss of generality,
we suppose $$q_1=0=c^2_{+}c^3_{-}-c^3_{+}c^2_{-}.$$
From this and \eqref{883} then follows that $l_2=l_3$, therefore $D_2=D_3$. We have thus proven that the
Poisson-Lie-WZ deformation is compatible with integrability  only if  at least  two of the parameters $D_1,D_2,D_3$  are  equal to each other. This is  different situation with respect to the Poisson-Lie deformation which could be applied in the integrable way to the $\si$-model with arbitrary values of the parameters $D_1,D_2,D_3$.

 \section{Conclusions and outlook} 
 We have reformulated the elliptic deformation of the principal chiral model as well as its Poisson-Lie deformation in terms of the $\E$-models which leads to a simple rederivation of the Lax pairs and RG flows of the models. Then we have constructed the Poisson-Lie-WZ deformation and determined the  values of the
 coupling constants for which the integrability is guaranteed. 

 As for the outlook, we believe that the $\E$-model approach to the determination of the  Lax pairs may be useful for construction of multiparametric deformations
 of the principal chiral (or Yang-Baxter) model on arbitrary simple Lie group which could go beyond the framework considered in \cite{LW24}.

 \end{document}